\documentclass[prl,10pt,aps,superscriptaddress,footinbib,twocolumn,floatfix,amssymb,amsmath,amsfonts]{revtex4-2}

\usepackage{epsfig}
\usepackage{slashed}

\usepackage{color}
\usepackage{xcolor}
\usepackage{graphicx}

\definecolor{limegreen}{RGB}{154,205,0}
\definecolor{brickred}{rgb}{0.8, 0.25, 0.33}
\definecolor{amethyst}{rgb}{0.6, 0.4, 0.8}
\definecolor{azure}{rgb}{0.0, 0.5, 1.0}
\definecolor{awesome}{rgb}{1.0, 0.13, 0.32}
\definecolor{purple}{rgb}{0.8,0,0.6}

\usepackage[normalem]{ulem}

\newcommand{\beqn}{\begin{eqnarray}}
\newcommand{\eeqn}{\end{eqnarray}}
\newcommand{\beqs}{\begin{subequations}}
\newcommand{\eeqs}{\end{subequations}\\[-2mm]\noindent}
\newcommand{\eq}[1]{(\ref{#1})}

\newcommand{\mr}{{\mathrm{mr}}}

\newcommand{\re}{\mathrm{Re}}
\newcommand{\rl}{\mathrm{Rl}}
\newcommand{\st}{\mathrm{St}}

\newcommand{\Karman}{K\'arm\'an\ }

\newcommand{\bs}{\boldsymbol}

\allowdisplaybreaks

\begin{document}

\title{Preturbulence in momentum-relaxing Navier-Stokes hydrodynamics}

\author{Vladimir A. Goy }
\affiliation{Pacific Quantum Center, Far Eastern Federal University, Sukhanova 8, Vladivostok, 690950, Russia}

\author{Uri Vool}
\affiliation{Max-Planck-Institut für Chemische Physik Fester Stoffe, Dresden, Germany}

\author{Maxim N. Chernodub}
\affiliation{Institut Denis Poisson UMR 7013, Universit\'e de Tours, 37200 France}

\begin{abstract}
In a hydrodynamic regime of electronic fluids, the electron scattering on impurities and phonons inhibits the development of oscillating preturbulent phenomena (such as the \Karman vortex shedding) and enforces a laminar flow. Working in a local generalization of the Navier-Stokes hydrodynamics of incompressible two-dimensional fluids, we show that the critical relaxation time separating these two regimes is described by a surprisingly simple fractional power function of the Reynolds number. The critical exponent, $\alpha \simeq 4/3$, sets particular experimental conditions for the observation of preturbulent effects in electronic fluids.
\end{abstract}

\date{\today}

\maketitle

%%%%%%%%%%%%%%%%%%%%%%%%%%%%%%%%%%%%%%%%%%%%%%%%%%

\paragraph{\bf Introduction.}

In certain solid-state systems, electronic transport exhibits collective hydrodynamic behavior, which appears to be in sharp contrast with the expected diffusive drift of a Fermi liquid~\cite{polini2019viscous}. Instead, the electron flow resembles a fluid when the momentum-conserving electron-electron scatterings events occur much faster than the timescale of the momentum-relaxing interactions of electrons with phonons, impurities, and the boundaries of the system~\cite{lucas2018hydrodynamics,Narozhny:2022ncn}.

The hydrodynamic behavior of electrons reveals itself via various transport properties in conductors of finite geometries. For example, the Gurzhi effect, which emerges in clean conducting channels, discriminates between ballistic transport restrained by the scattering of electrons with the boundaries of the channel and laminar flow of electronic fluid, dominated by the momentum-conserving electron-electron scattering events~\cite{ref_Gurzhi_1963}. The robust signatures of the electron hydrodynamic flow have been found in various compounds via its imprints on the thermal and electric transport~\cite{ref_Molenkamp,ref_hydro_PdCoO2,ref_hydro_WP2,ref_WTe2_flow}.

The exceptional mobility of electrons in ultraclean graphene offers another exciting possibility to investigate their hydrodynamics~\cite{muller2009graphene}. Due to the viscosity of the electronic flow~\cite{govorov2004hydrodynamic,torre2015nonlocal,pellegrino2016electron,bandurin2018fluidity} a localized injection of the electronic current can create whirlpools in the electron flow near the contacts. The emerging electronic vortices were predicted to generate a viscous backflow~\cite{ref_Levitov,Lucas_2018} which leads to experimentally observable negative resistance~\cite{ref_graphene_flow_1} as well as a strong violation of the Wiedemann-Franz law~\cite{ref_graphene_wiedemann}. The viscosity of electronic flow reveals itself in hydrodynamic conductance in narrow constrictions~\cite{ref_graphene_flow_3}, which allows for the visual discrimination\footnote{Notice that visually compelling hydrodynamic-like response may also emerge in the non-hydrodynamic ballistic regime~\cite{shytov2018particle}, via surface scattering (``para-hydrodynamics'')~\cite{AharonSteinberg2022}, and in diffusive two-component systems (``pseudo-hydrodynamics'')~\cite{Choi2022}.} of the high-temperature diffusive Ohmic current from the low-temperature laminar Poiseuille flow via spatial imagining of the electronic current profiles~\cite{ref_graphene_flow_2,ref_graphene_flow_4,ref_graphene_flow_5}.

The experimental efforts are mostly concentrated in the laminar hydrodynamical regime characterized by low values ($\re \ll 1$) of the Reynolds number:
\beqn
\re = \frac{U L}{\nu}\,,
\label{eq_Re}
\eeqn
set by the characteristic linear dimension $L$ of the hydrodynamic problem, the flow velocity $U$, and its kinematic (shear) viscosity $\nu$.

At higher Reynolds numbers, typically $\re \simeq 10 \dots 10^3$, electronic transport may exhibit ``preturbulent'' phenomena~\cite{muller2009graphene}, which may experimentally be observed via fluctuations of the electronic current past constrictions and localized impurities in the flow channel~\cite{mendoza2011preturbulent,Gabbana2018}. In ordinary fluid dynamics, the preturbulent regime is manifested via a periodic detachment of hydrodynamic vortices, which form the so-called \Karman vortex street~\cite{Von_Karman2004-lu,Falkovich2018}. In condensed matter context, the formation of the \Karman vortices has been theoretically predicted~\cite{superfluidKarmanPrediction} and experimentally observed~\cite{superfluidKarmanObservation} in the Bose-Einstein condensates of cold atoms.

Contrary to the flow of ordinary fluids, which experience momentum-relaxing interactions only with the channel boundaries and macroscopic obstacles in the flow, the electronic fluids can also lose their momentum at a microscopic level via inelastic scattering of electrons on thermal phonons and impurities of a crystal lattice. These momentum-relaxing events can substantially inhibit the development of the preturbulence regime.

The aim of our paper is to investigate the preturbulent regime in two-dimensional incompressible fluids in the presence of a uniform background of momentum-relaxing events. We consider Navier-Stokes (NS) fluids extended with a simple momentum-relaxing term and study its effect in the realm of electronic flows in graphene. Due to the minimal and model-independent nature of the extension, this simple model can serve as a universal benchmark case of NS fluids with dissipation to be compared with more elaborate models~\cite{mendoza2011preturbulent,Gabbana2018}.

\vskip 1mm

\paragraph{\bf Navier-Stokes hydrodynamics with relaxation.}

A one-component electronic fluid is characterized by the electronic density $n$ of the charge carriers with the mass $m$, the particle current density $\bs j = n {\bs u}$ and the local velocity ${\bs u} = {\bs u}({\bs r})$. We consider a fluid propagating slower than plasmonic velocities which possesses incompressible flow, ${\bs \nabla} \cdot {\bs u} = 0$~\cite{bandurin2018fluidity,ref_Levitov}. Its hydrodynamic regime is described by the extended Navier-Stokes equation:
\begin{equation}
\biggl( \frac{\partial }{\partial t} + {\bs u} \cdot {\bs \nabla} \biggr) {\bs u}
- \nu \Delta {\bs u} = - \frac{1}{m n} {\bs \nabla} P - \frac{e}{m} {\bs E} - \frac{1}{\tau_{\mr} } {\bs u}\,,  
\label{eq_Navier_Stokes_1}
\end{equation}
which incorporates forces experienced by an element of the fluid including (i) internal shear stress due to velocity gradients with the magnitude set by the kinematic shear viscosity $\nu$, (ii) the pressure-gradient ${\bs \nabla} P$, (iii) the background electrostatic field ${\bs E}$, and (iv) friction due to electron scattering on phonons and crystal defects.

The parameterization of the friction term by the single quantity, the scattering time $\tau_{\mr}$, turns out to be remarkably successful in explaining experimental results in the linear-response regime~\cite{ref_graphene_flow_1,ref_graphene_flow_3,bandurin2018fluidity,Braem2018}.

A steady laminar flow can be described by the simplified time-independent NS equation~\eq{eq_Navier_Stokes_1}, with the material derivative and pressure-gradient terms neglected~\cite{ref_graphene_flow_1}: $\sigma_0 {\bs E} ({\bs r}) - e L_{\mathrm{Gu}}^2 \Delta {\bs j}({\bs r}) + e {\bs j}({\bs r}) = 0$. The spatial hydrodynamic length scale is set by the (diffusive) Gurzhi length 
\begin{align}
    L_{\mathrm{Gu}} = \sqrt{{\nu} \tau_{\mathrm{mr}}}\,.
\label{eq_Gurzhi}
\end{align}
which, for example, determines the spatial size of the electronic whirlpools in graphene~\cite{ref_Levitov}. At vanishing viscosity, $\nu \to 0$, or at high momentum-relaxing collision rate, $\tau_{\mathrm{mr}} \to 0$, the hydrodynamic length vanishes, $L_{\mathrm{Gu}} \to 0$, and the NS equation reduces to the usual Ohmic drift of electric current, ${\bs J} \equiv - e {\bs j} = \sigma_0 {\bs E}$, set by the Drude-like conductivity $\sigma_0 = n_0 e^2  \tau_{\mathrm{mr}}/m$.

\vskip 1mm
\paragraph{\bf Scales.}

The vortex shedding occurs in the non-linear regime of the NS equation~\eq{eq_Navier_Stokes_1} which can be conveniently rewritten in the following dimensionless form:
\beqn
\re \left(\frac{\partial}{\partial t_0} + {\bs u}_0 \cdot {\bs \nabla}_0 \right) {\bs u}_0 - \Delta_0 {\bs u}_0 = - \rl\, {\bs u}_0 + {\bs f}_0\,, 
\quad
\label{eq_NS_rescaled}
\eeqn
where dimensionless quantities carry the subscript~``0'': ${\bs x} = L {\bs x}_0$, $t = T_L t_0$, ${\bs u} =  U {\bs u}_0$, including the dimensionless force ${\bs f}_0 = - T_L ({\bs \nabla}_0 P)/(m n \nu) - e L T_L  {\bs E}/(m \nu)$. The hydrodynamics of fluid~\eq{eq_NS_rescaled} is characterized by two dimensionful time scales: the kinematic time $T_L = L/U$, which sets, for example, the scale for the shedding frequency of the \Karman vortices [see Eq.~\eq{eq_f_St}], and the kinematic relaxation time scale $\tau_{L} = L^2/\nu$, which determines their attenuation. In Eq.~\eq{eq_NS_rescaled}, we also introduced the dimensionless momentum-relaxation number,
\begin{align}
    \rl = \frac{\tau_{L}}{\tau_{\mr}} = \frac{L^2}{\nu \tau_\mr} = \frac{L^2}{L_{\mathrm{Gu}}^2}\,.
    \label{eq_Rl}
\end{align}
given by the ratio of kinematic, $\tau_L$, and actual, $\tau_\mr$, momentum-relaxation times. This parameter separates regimes of fast ($\tau_L \gg \tau_\mr$ or $\rl \gg 1$) and slow ($\tau_L \ll \tau_L$ or $\rl \ll 1$) momentum relaxation. Thus, the hydrodynamics can be fully described by only two dimensionless numbers: The Reynolds number $\re$~\eq{eq_Re} and the momentum-relaxation number $\rl$~\eq{eq_Rl}.

\vskip 1mm
\paragraph{\bf The critical line of preturbulence.}

We extensively simulated the NS fluid~\eq{eq_NS_rescaled} to elucidate the effects of the momentum relaxation term on the onset of preturbulence. We consider a circular impurity of the diameter $L$ placed in the center of a wide straight channel of the large width $W = 10 L$ and use no-slip boundary conditions with vanishing velocity at the boundaries. 

In the absence of momentum relaxation, $\tau_{\mr} \to \infty$, the flow becomes unstable at high Reynolds numbers~\eq{eq_Re}, $\re > \re_c$, where the critical value $\re_c$ depends on the geometry of the hydrodynamic system. For a disk-shaped two-dimensional obstacle in a channel of infinite width, the \Karman vortices start to form if the Reynolds number exceeds the critical value $\re^{(\infty)}_c \simeq 47$~\cite{Jackson1987}. In our geometry with a finite-size channel, the critical Reynolds number in the relaxation-free ($\tau_\mr = \infty$) regime is:
\beqn
    \re_c = 52.4(5)\,, \qquad ({\textrm{at}\ }\tau_{\mr} \to \infty)\,.
    \label{eq_Re_c}
\eeqn

We found numerically that a finite momentum relaxation time $\tau_{\mr}$  drastically affects the value of the critical Reynolds number by inhibiting the formation of the \Karman vortices in agreement with Ref.~\cite{Gabbana2018}. The phase diagram of the preturbulence in the $(\re,\rl)$ plane is shown in Fig.~\ref{fig_strength} (the amplitude of the flow velocity is explored in more detail in Supplementary Figure~\ref{fig_relative_amplitude}).

\begin{figure}[!htb]
\begin{center}
\includegraphics[width=0.95\linewidth, clip=true]{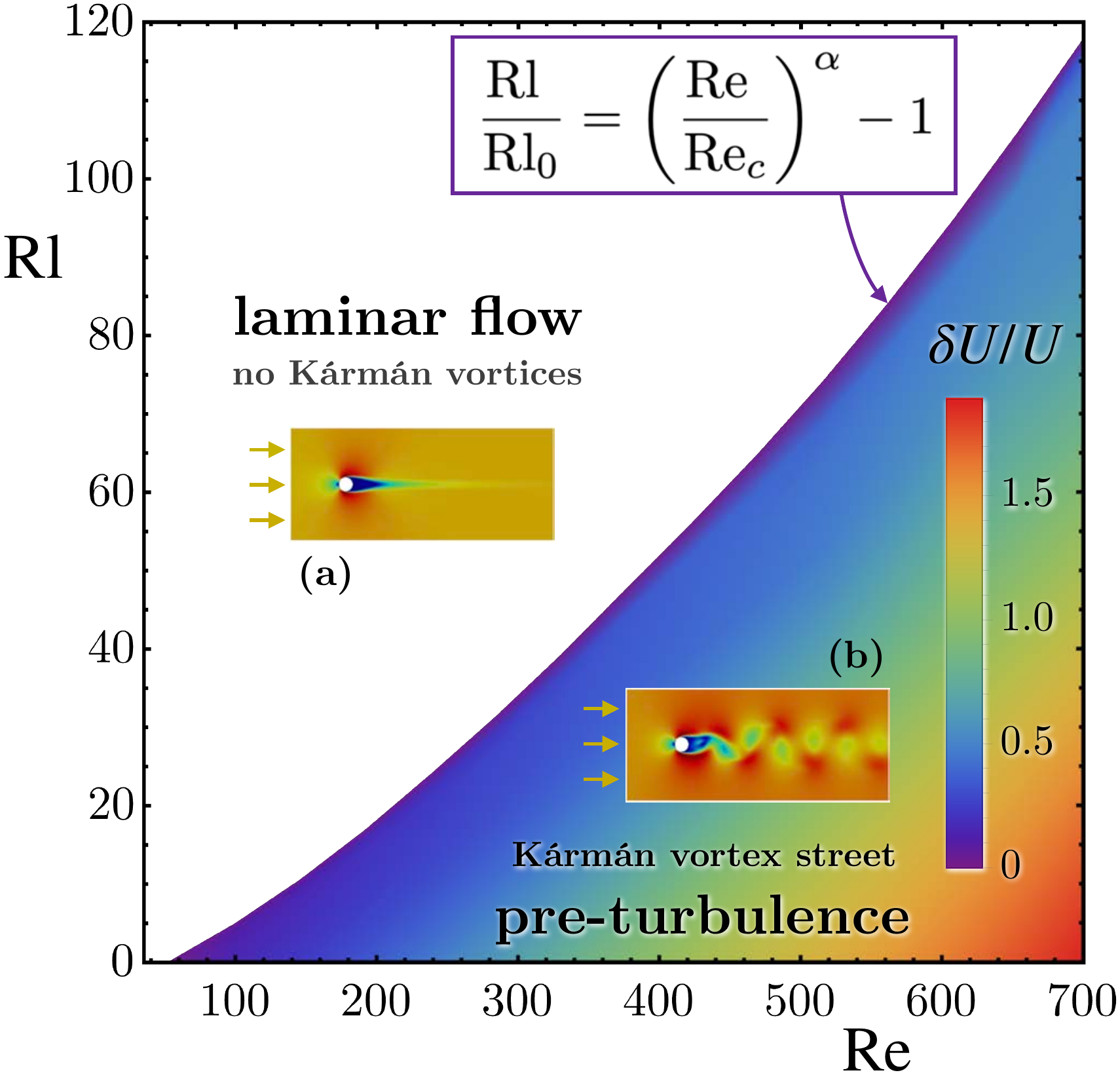} 
\end{center}
\vskip -5mm
\caption{Phase diagram of the vortex formation in the plane of the Reynolds number $\re$, Eq.~\eq{eq_Re}, and the momentum-relaxation number $\rl$, Eq.~\eq{eq_Rl}, past a circular impurity of a diameter~$L$ in the channel of the width $W = 10 L$. The critical line, described by the phenomenological fractional power law~\eq{eq_tau_c} with parameters~\eq{eq_fit_parameters}, separates the laminar flow (the white region above the transition line at low $\tau_{\mr}$, high $\rl$) and the preturbulent region with higher $\tau_{\mr}$ (lower $\rl$), where the flow produces the \Karman vortices. The examples of the laminar flow and the oscillatory \Karman flow are shown in the insets (a) and (b), respectively. The color displays the magnitude of velocity oscillations at the point $r = L$ in the middle of the channel past the obstacle.}
\label{fig_strength}
\end{figure}

The phase diagram possesses a laminar phase (at $\tau_\mr < \tau_{\mr,c}$) and the pre-turbulent \Karman phase (at $\tau_\mr > \tau_{\mr,c}$). We established that the critical line that separates these phases is described by the fascinatingly simple fractional power law:
\beqn
\frac{\rl\ }{\rl_0} = \left( \frac{\re\ }{\re_c} \right)^\alpha - 1, 
\qquad 
\re \geqslant \re_c\,,
\label{eq_tau_c} 
\eeqn
where the best-fit parameters are as follows:
\beqn
\rl_0 = 3.63(7)\,, 
\qquad
\alpha = 1.348(2) \simeq 4/3\,,
\label{eq_fit_parameters}
\eeqn
and the critical Reynolds number is given in Eq.~\eq{eq_Re_c}. The excellence of the fit of the numerical data by the phenomenological scaling~\eq{eq_tau_c} is demonstrated in Fig.~\ref{fig_fitting}.

The relative amplitude of the velocity oscillations, $\delta U/U$, created by the vortices produced past the impurity, takes its weakest values near the critical line~\eq{eq_tau_c}. The amplitude gets substantially enhanced as either the Reynolds number Re or the relaxation time $\tau_\mr$ increases.

In a nondissipative limit, when the relaxation time is long, $\tau_{\mr} \gg \tau_L$ ($\rl \ll 1$), the vortices start to form closer to the lowest critical Reynolds number~\eq{eq_Re_c}. At higher Reynolds values, the preturbulent vortex regime sets in at a shorter momentum-relaxation time (for example, according to Fig.~\ref{fig_strength}, the critical relaxation time is $\tau_{\mr,c} \simeq \tau_L/60$ or $\rl_c \simeq 60$ at $\re = 400$).

Equations~\eq{eq_Re}, \eq{eq_Rl} and \eq{eq_tau_c} imply that the \Karman vortex phase is realized at the initial flow velocities higher than the critical value $u > u_c$:
\beqn
u_c(\tau_{\mr}) = \re_c \, \frac{\nu}{L} \left(\frac{1}{\rl_0} \, \frac{L^2}{\nu \tau_{\mr}} + 1 \right)^{1/\alpha}\,.
\label{eq_u_c}
\eeqn

In the absence of momentum relaxation, $\tau_{\mr} \to \infty$, the critical line~\eq{eq_u_c} is reduced to $u_c(\tau_{\mr}) = \re_c \, \frac{\nu}{L}$ which defines the critical point~\eq{eq_Re_c} $\re = \re_c$, as expected. In the opposite, fast-relaxation limit, $\tau_{\mr} \ll \tau_L$ (at $\rl \gg 1$), one gets the following asymptotic scaling regime:
\beqn
u_c(\tau_{\mr}) = u_0 \, \nu^{\frac{\alpha - 1}{\alpha}} L^{\frac{2 - \alpha}{\alpha}} \tau^{-\frac{1}{\alpha}}_{{\mr}} 
\quad\ 
\mbox{for} \quad \tau_{\mr} \ll L^2/\nu, \quad
\label{eq_u_c_lim}
\eeqn
with the dimensionless $u_0 = \re_c \, \rl_0^{-1/\alpha} = 20.1(2)$.

\begin{figure*}[!htb]
\begin{center}
\includegraphics[width=0.95\linewidth, clip=true]{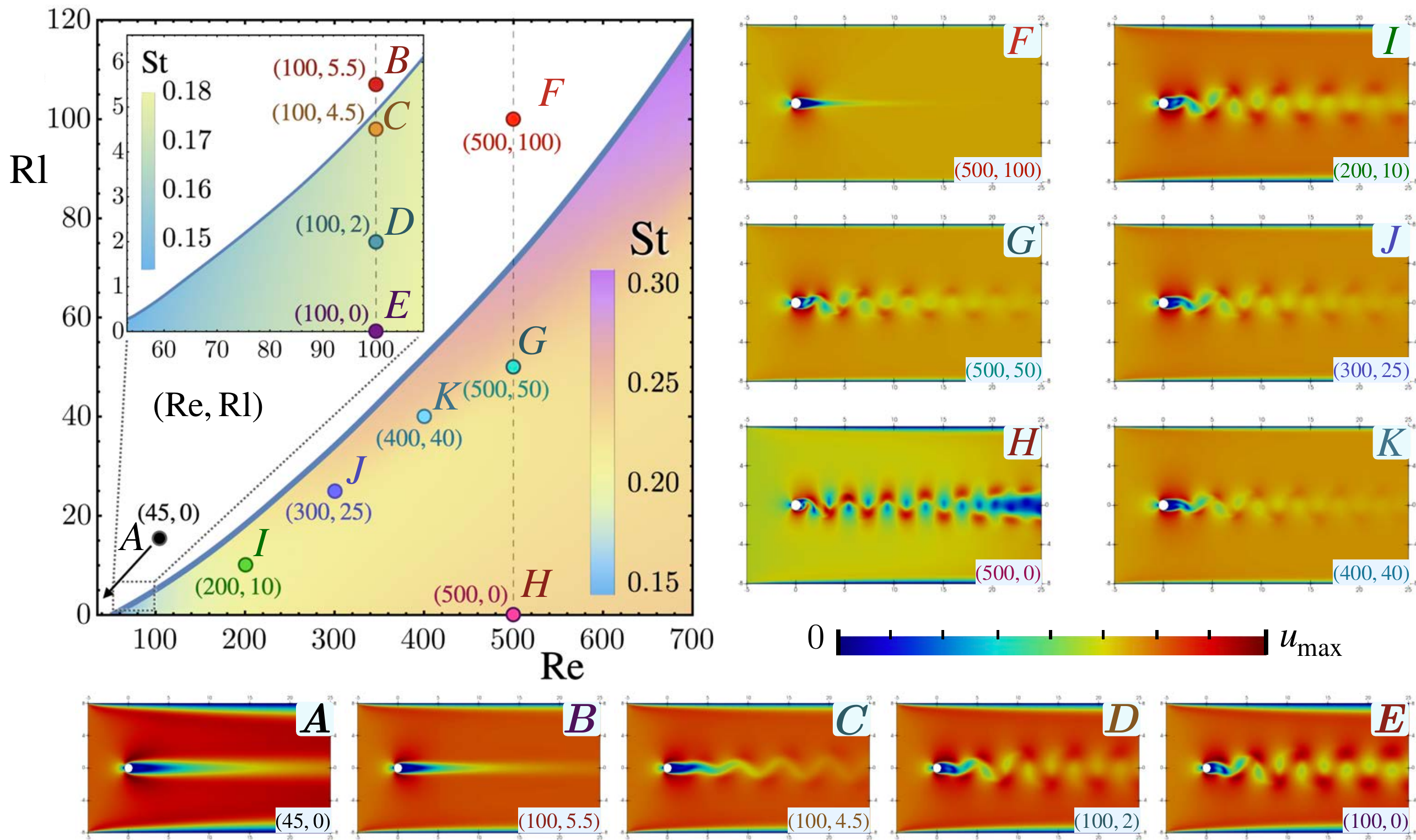} 
\end{center}
\caption{
The Strouhal number St, which determines the frequency of the vortex oscillations~\eq{eq_f_St}, is shown in the (Re,\,Rl) plane of the phase diagram of Fig.~\ref{fig_strength} along with the examples of the 
\Karman vortex streets and \Karman tadpoles at the points $A, B, \dots, K$ (the color marks the absolute value of the velocity flow). The inset shows a low-Re corner of the diagram.
}
\label{fig_frequency}
\end{figure*}

\vskip 1mm
\paragraph{\bf Shedding frequency and momentum relaxation.}

The vortex production is characterized by a periodic pattern of the unsteady flow of the vortex shedding. In ordinary fluids, the main vortex shedding frequency,
\beqn
f = \st \frac{U}{L} \equiv \frac{\st}{T_L} \,,
\label{eq_f_St}
\eeqn
is expressed via the dimensionless Strouhal number ``St''~\cite{Falkovich2018}. The Strouhal number, which can be computed numerically, depends on the geometry of the obstacle.

The typical frequency of the vortex shedding~\eq{eq_f_St} is set by the kinematic timescale $T_L$ which depends on the velocity of the fluid and the size of the obstacle. For a disk-shaped impurity in the absence of the momentum relaxation, $\tau_{\mr} = \infty$, the Strouhal number takes values around $\st \simeq 0.2$ in a wide range of values of the Reynolds number~\cite{Chen1985}. In a momentum-nonconserving fluid, the finite relaxation time scale $\tau_\mr$ implicitly affects the vortex shedding frequency~\eq{eq_f_St}. Contrary to the soft equation of state~\cite{Gabbana2018}, the fluid oscillations in our model~\eq{eq_NS_rescaled} feature a single production frequency~\eq{eq_f_St}.

Figure~\ref{fig_frequency} shows the Strouhal number calculated numerically across the whole phase diagram. In the low-Re corner of the diagram, corresponding to slow relaxation with high $\tau_{\mathrm{mr}}$, the Strouhal number takes the lowest value around $\st \simeq 0.15$. In contrast, in the opposite high-Re and low-$\tau_{\mr}$ corner, the oscillations of the vortex flow are characterized by a twice higher Strouhal number, $\st \simeq 0.3$. Finally, in the rest of the phase diagram, the oscillations are close to the standard value $\st \simeq 0.2$.

\vskip 1mm
\paragraph{\bf \Karman vortices and \Karman tadpoles.}

Figure~\ref{fig_frequency} also illustrates qualitative features of the fluid flow in various points $A$-$K$ of the phase diagram. The momentum-relaxing laminar flow above the critical line (at $\re > \re_c$ and $\tau_\mr \neq 0$ with $\tau_\mr < \tau_{\mr,c}$, point~$A$ in Fig.~\ref{fig_frequency}), appears to be qualitatively similar to the ordinary laminar phase below the critical Reynolds number ($\re < \re_c$ at $\tau_\mr = \infty$, point~$B$). The only difference between these phases appears, expectedly, in a faster dissolution of the long-distance laminar tail at $\tau_\mr \neq 0$ (well pronounced at point $F$ at a smaller $\tau_\mr$).

In the preturbulent phase, the momentum relaxation produces a visible qualitative effect on the structure of the vortex shedding. For example, at $\re = 100$, the vortices become individually non-distinguishable as they form a single oscillating trail which we dub the ``\Karman tadpole'', which is well-visible in point $C$. As the relaxation time $\tau_\mr$ increases, the tadpole breaks into individual vortices (point $D$), which become more pronounced as $\tau_\mr$ reaches the non-dissipative limit (point $E$). Thus, at $\re > \re_c$, an increasing relaxation time $\tau_\mr$ turns a laminar flow ($B$) into a \Karman tadpole ($C$), which then gets gradually transformed into the usual \Karman vortex street ($D$). The effect of the momentum relaxation on the vortex attenuation is also seen in the series of points $F$-$H$ and $I$-$K$ at higher Reynolds numbers.

\vskip 1mm
\paragraph{\bf Large holes.}

\begin{figure}[!htb]
\begin{center}
\includegraphics[width=0.9\linewidth, clip=true]{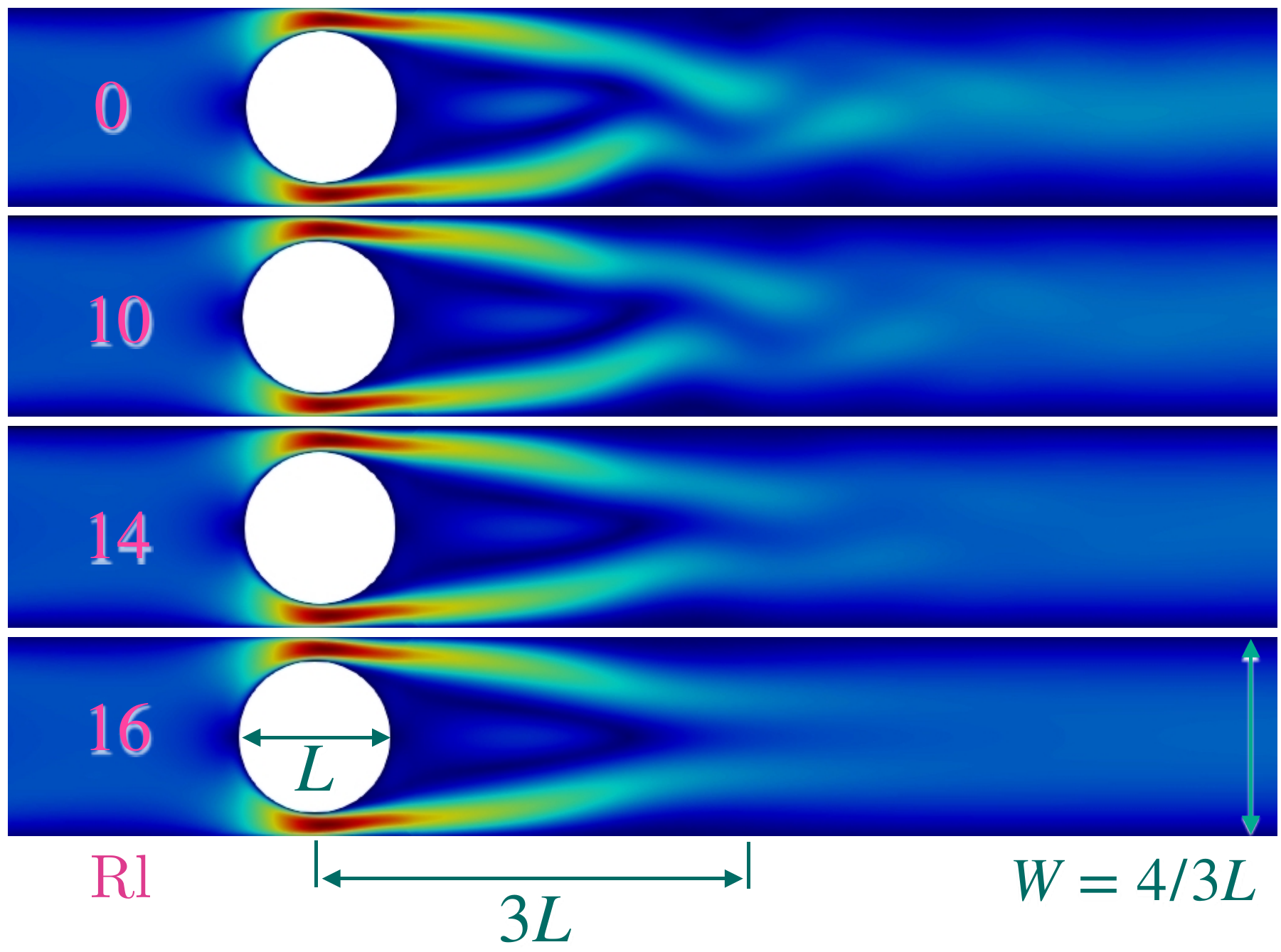} 
\end{center}
\vskip -3mm
\caption{Absolute value of the flow velocity around a large circular impurity of the diameter $L$ in a channel of the width $W = 4/3 L$ for various momentum-relaxation numbers $\rl$. $\re=100$, determined formally via the hole size~\eq{eq_Re}, is fixed for all maps, and we observe preturbulence with the critical value $\rl_c \approx 15$. The velocity color map follows Fig.~\ref{fig_frequency}.}
\label{fig_large_hole}
\end{figure}

A non-zero momentum relaxation also produces an attenuating effect for large holes, which may have important consequences relevant to the experimental observation of preturbulence. In Fig.~\ref{fig_large_hole}, we show the velocity flow around a large disk-shaped impurity for several momentum-relaxation times $\tau_\mr$. An impurity of diameter $L$ in a channel of slightly larger width $W = 4/3 L$ separates the fluid flow into two narrow constrictions. The constrictions inject the fluid with an enhanced velocity past the obstacle close to the walls of the channel, leading to the formation of the \Karman vortex street.

At a distance $r \simeq (5-6) L$ from the obstacle, the flow becomes laminar. A decreasing relaxation time inhibits the preturbulence effect, leading to a pure laminar flow of two joining sub-flows above the critical value $\rl \gtrsim 15$. Our results, illustrated in Fig.~\ref{fig_large_hole}, imply that the most favorable point for the observation of vortices is $r \simeq (2-3) L$ past the impurity. The relevant frequencies/amplitudes are of the same order as for small holes with important geometry-dependent corrections that are explored in more detail in Supplementary Fig.~\ref{fig_varying_diameters}.

\vskip 1mm
\paragraph{\bf Realistic experimental parameters and measurement techniques.}

With this quantitative understanding of the threshold for \Karman vortices, let us now examine if realistic parameters can be chosen that cross this threshold in an experimental system,  focusing on high quality low carrier-density conductors such as graphene sheets and GaAs quantum wells.
Initially, a large flow velocity~$U$ is required, and velocities of $U \approx (1-5) \times 10^5~\mathrm{m/s}$ have been previously achieved in GaAs~\cite{smith_velocity_1980,mokerov_drift_2009} and graphene~\cite{Meric2008,dorgan2010mobility,andersen_electron-phonon_2019}. The Reynolds number can be increased by constrictions of larger diameter~$L$, but this also increases the relaxation timescale~$\tau_L$, thus placing a larger constraint on the relaxation time~$\tau_\mr$. This trade-off is explored in detail in Supplementary Figure~\ref{fig_velocity}, which shows the required velocity $U$ and viscosity $\nu$ to cross the vortex-shedding threshold for various constriction diameters~$L$ and momentum relaxation times~$\tau_{\mr}$. 

As $\tau_{\mr}$ is currently experimentally limited to $1{-}10 \,\mathrm{ps}$~\cite{ref_Molenkamp,ref_graphene_flow_1}, a diameter length scale of $L \approx 1 \,\mu \mathrm{m}$ seems optimal. This experimentally favorable condition is achieved in the asymptotic scaling regime~\eq{eq_u_c_lim} which gives $u_c \propto \nu^{1/4} L^{1/2} \tau^{- 4/3}_{{\mr}}$ due to the relatively large~\eq{eq_fit_parameters} scaling exponent $\alpha > 1$. The observation of preturbulence requires then viscosity $\nu \approx 10^{-4} \mathrm{m^2/s}$. Such a value could be possible~\cite{ref_graphene_flow_1}, but is currently the most significant deviation of needed parameters from existing measured values, consistent with a previous analysis of preturbulence~\cite{Gabbana2018}. 
Improvements in relaxation time can allow for larger constriction sizes, thus reducing the viscosity constraint.

As the effect of \Karman vortices is local to the area past the impurity, the strong effects will be difficult to measure in regular transport measurements but will require local probes such as electric potential~\cite{ref_graphene_flow_2}, scanning SQUID magnetometry~\cite{AharonSteinberg2022}, or magnetometry with nitrogen-vacancy (NV) centers~\cite{ref_graphene_flow_4,ref_graphene_flow_5,ref_WTe2_flow}. 
The latter is particularly advantageous due to the high sensitivity of the NV to microwave radiation at frequencies corresponding to its spin transitions for a given applied magnetic field. 
Thus, in addition to the measurement of the DC current shape, the oscillations themselves can be locally observed, with the ability to tune their frequency in-situ by varying the applied current and thus the velocity $U$. 
The oscillation frequency~\eq{eq_f_St} of the vortices for this typical geometry is $f\approx 10 \,\mathrm{GHz}$, which is conveniently obtainable for NV centers with the applied field. The observation of a local microwave signal (when only a DC signal is applied) that is tunable with the applied current will be a smoking-gun signature of preturbulent electron hydrodynamics.

\vskip 1mm
\paragraph{\bf Conclusions.}

Our paper explored the turbulent-dissipative regime in two-dimensional incompressible electronic fluids by taking the most straightforward local extension of the NS equation with the local momentum-relaxation term~\eq{eq_NS_rescaled} arising from electron scattering on impurities and phonons. We suggested characterizing the relaxation rate in terms of the new momentum relaxation number Rl~\eq{eq_Rl}. We found the critical relaxation number Rl that discriminates the laminar flow from the pre-turbulent regime, is described by a surprisingly simple fractional power law~\eq{eq_tau_c} of the Reynolds number Re with the exponent $\alpha \simeq 4/3$ which intriguingly matches the Yaglom turbulence exponent for dissipation rates of conserved
quantities~\cite{yaglom1949local,monin2013statistical}. 

Close to the critical transition line, the street of shredded \Karman vortices collapses to a new dynamical structure, a single vibrating \Karman tadpole. In the local relaxation model~\eq{eq_NS_rescaled}, the oscillations are characterized by a single global frequency that does not depend on the distance to the impurity. 

The novel preturbulent-dissipative regime accessible with electron hydrodynamics shows rich behavior that can be effectively modeled by two dimensionless parameters. This simple model will serve as a benchmark for the exploration of this regime and a useful tool to motivate, guide, and interpret future experimental studies.

\vskip 1mm
\begin{acknowledgments}
\paragraph{\bf Acknowledgments.}
The work of VAG was supported by Grant No. 0657-2020-0015 of the Ministry of Science and Higher Education of Russia.
\end{acknowledgments}

%%%%%%%%%%%%%%%%%%%%%%%%%%%%%%%%%%%%%%%%%%%%%%%%%%

%\bibliographystyle{apsrev4-1}

%\bibliography{KarmanRefs}

%apsrev4-2.bst 2019-01-14 (MD) hand-edited version of apsrev4-1.bst
%Control: key (0)
%Control: author (8) initials jnrlst
%Control: editor formatted (1) identically to author
%Control: production of article title (0) allowed
%Control: page (0) single
%Control: year (1) truncated
%Control: production of eprint (0) enabled
%

%%%%%%%%%%%%%%%%%%%%%%%%%%%%%%%%%%%%%%%%%%%%%%%%%%

\clearpage

\renewcommand\thefigure{S\arabic{figure}} 
\setcounter{figure}{0}

\renewcommand\theequation{S\arabic{equation}} 
\setcounter{equation}{11}

\renewcommand{\thesection}{S\arabic{section}}
\setcounter{section}{0}

\renewcommand{\thepage}{S\arabic{page}}
\setcounter{page}{1}

\onecolumngrid

\centerline{\Large{Supplementary Information}}
\vskip 5mm
\centerline{\bf \large{Preturbulence in momentum-relaxing Navier-Stokes hydrodynamics}}
\vskip 2mm
\centerline{by Vladimir A. Goy, Uri Vool, and Maxim N.~Chernodub}
\vskip 2mm

\appendix
\section{A. Amplitude of velocity fluctuations}

\begin{figure*}[!htb]
\begin{center}
\includegraphics[width=130mm, clip=true]{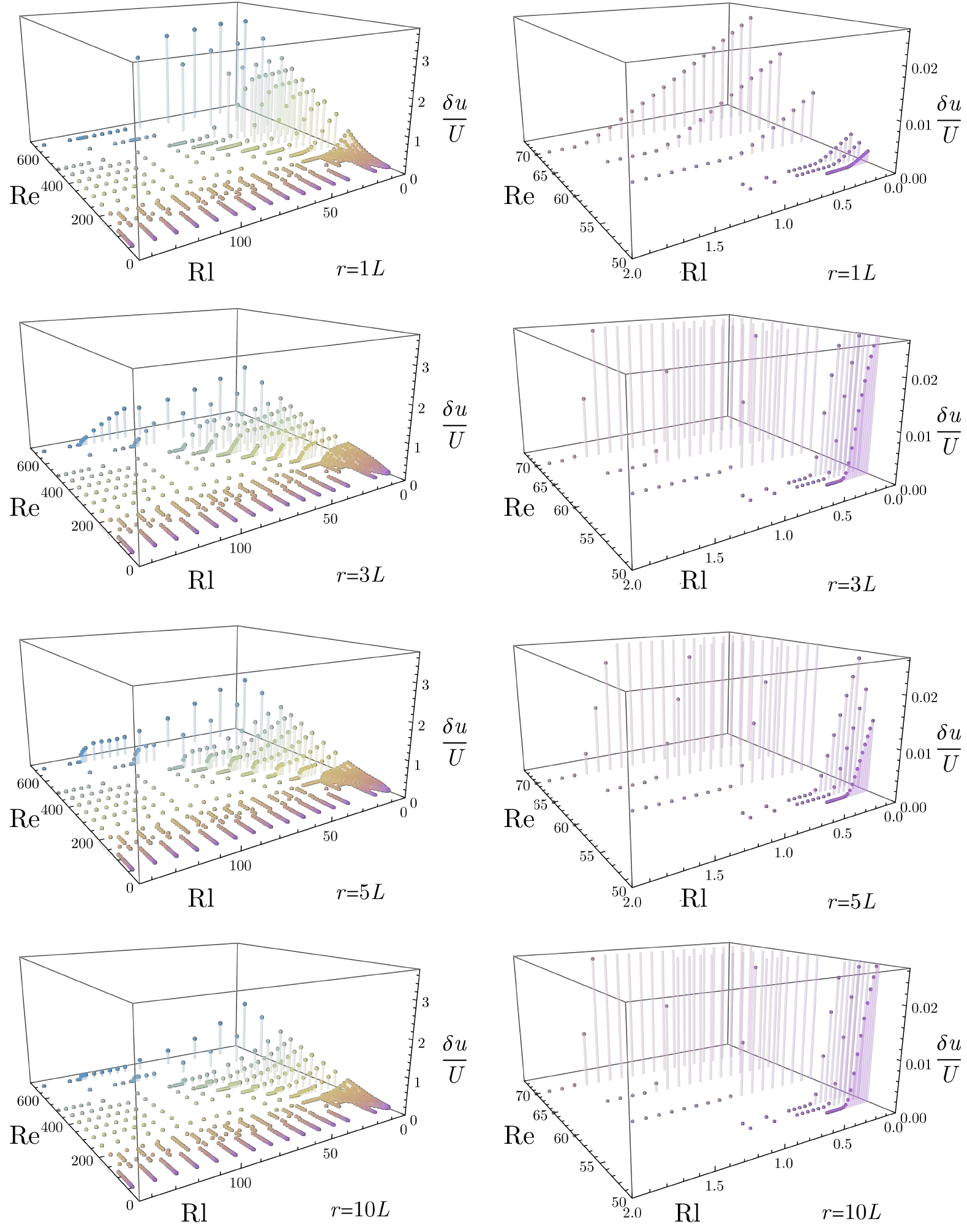}
\end{center}
\caption{The relative amplitude of the velocity fluctuations of the flow, $\delta u/U$ (with the initial flow velocity $U$), calculated at the distances $r = L, 3 L, 5 L, 10 L$ past the hole of the diameter $L$ in the middle of the channel as the function of the Reynolds number Re and the momentum-relaxation number $\rl$. The right column represents a zoom in on the small-Re corner of the corresponding plot at the left column.}
\label{fig_relative_amplitude}
\end{figure*}

\section{B. The best fit of the transition line.}

\begin{figure}[!htb]
\begin{center}
\includegraphics[width=80mm, clip=true]{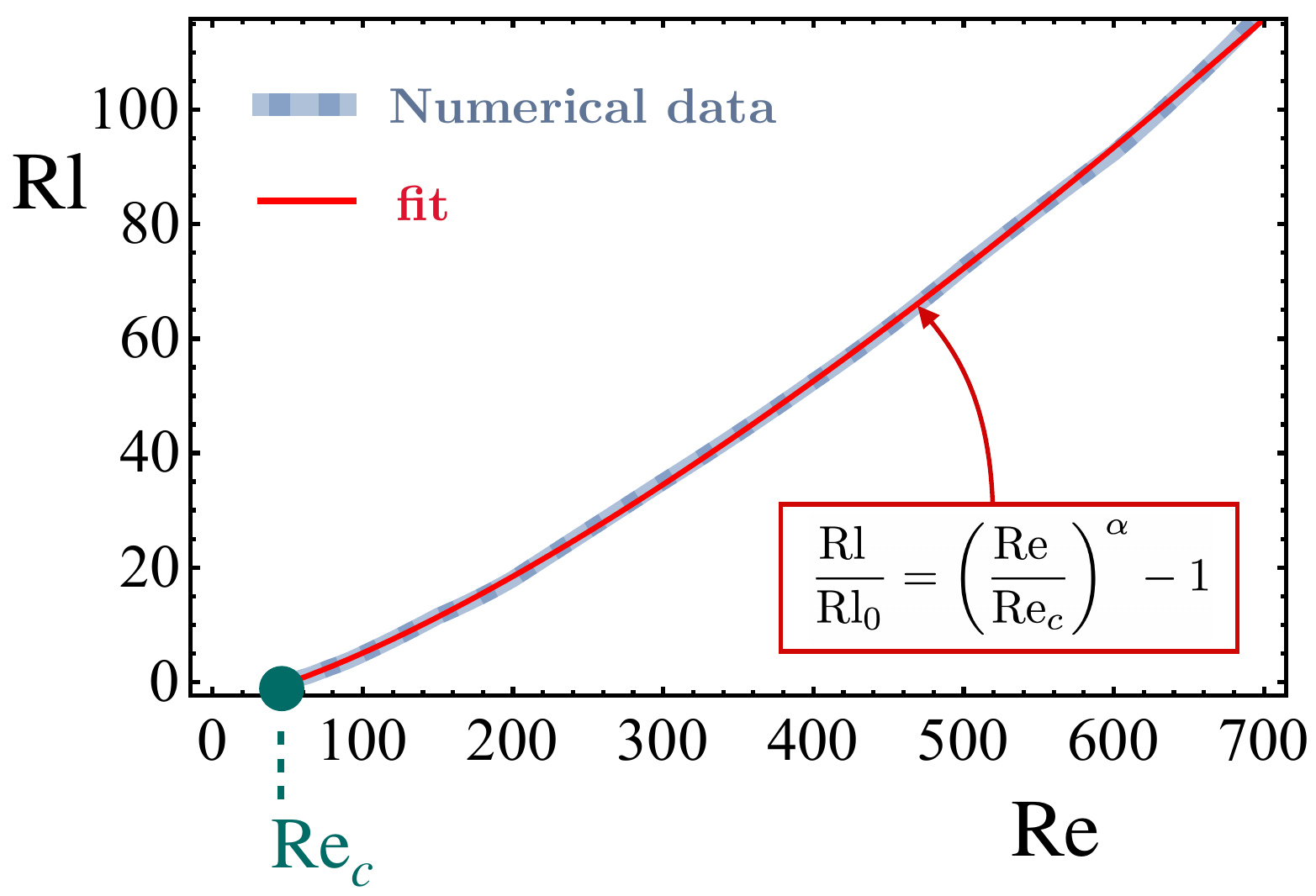} 
\end{center}
\caption{The best fit of the transition line which separates, in the plane of the Reynolds parameter $\re$ and the relaxation parameter $\rl$, the regions of the preturbulent and laminar phases in Fig.~\ref{fig_strength} and Fig.~\ref{fig_frequency}.
The fit is performed by the fractional power function~\eq{eq_tau_c} with the best-fit parameters given in Eq.~\eq{eq_fit_parameters}.}
\label{fig_fitting}
\end{figure}

\section{C. The transition line with physical parameters for electronic fluids}

\begin{figure*}[!htb]
\begin{center}
\begin{tabular}{ccc}
\includegraphics[width=0.33\linewidth, clip=true]{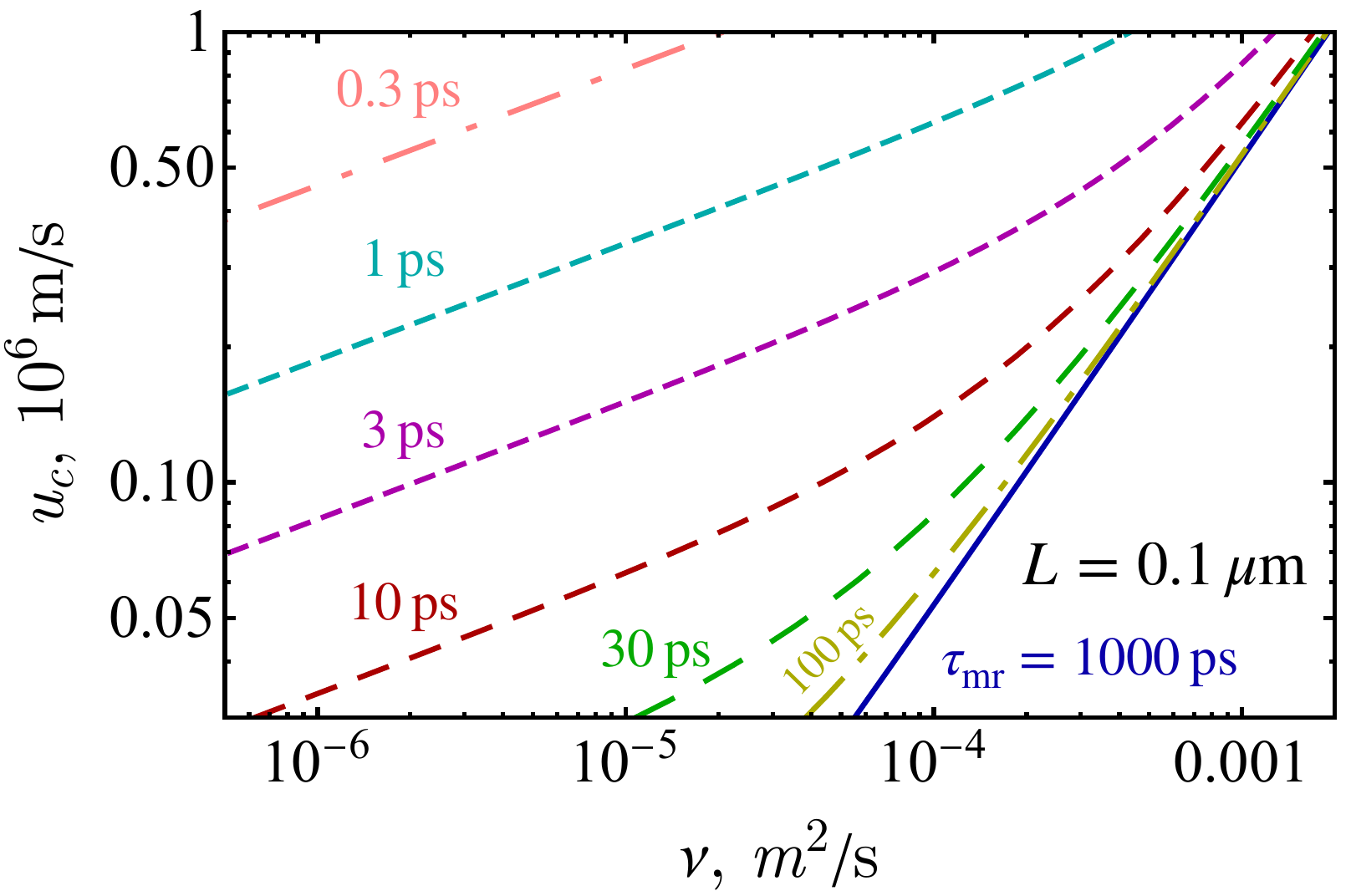} &
\includegraphics[width=0.33\linewidth, clip=true]{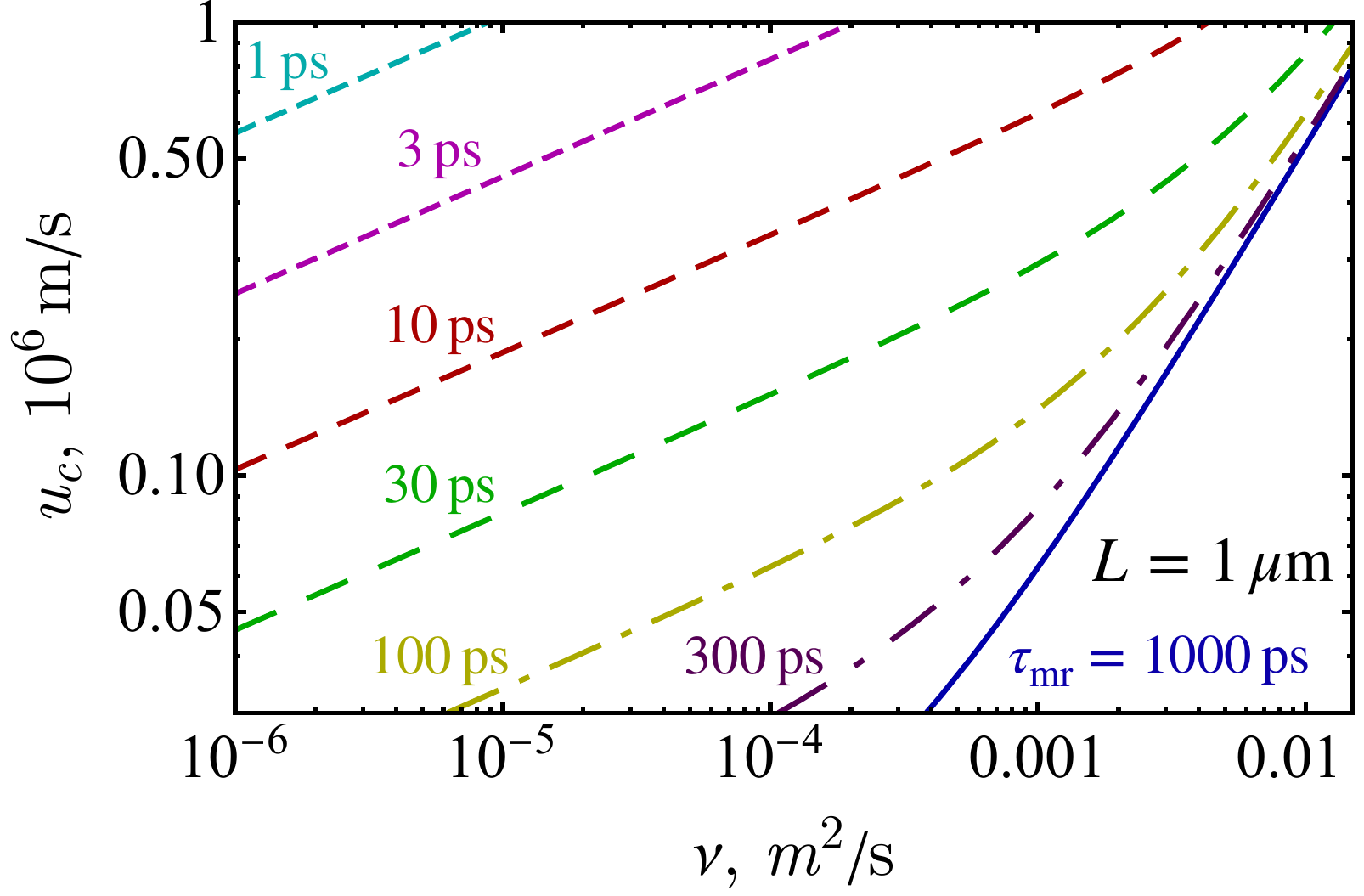} & 
\includegraphics[width=0.33\linewidth, clip=true]{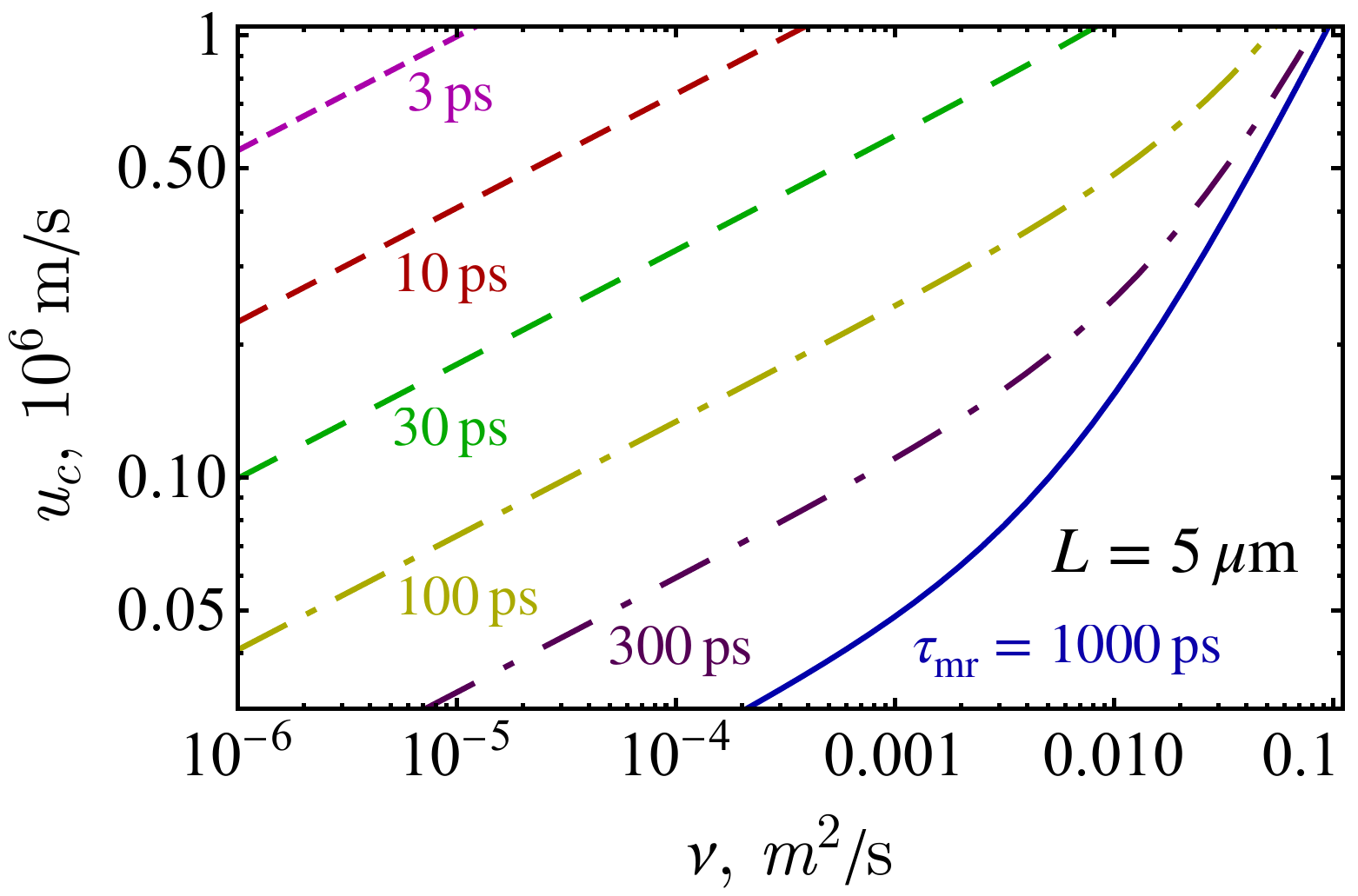} \\
\includegraphics[width=0.33\linewidth, clip=true]{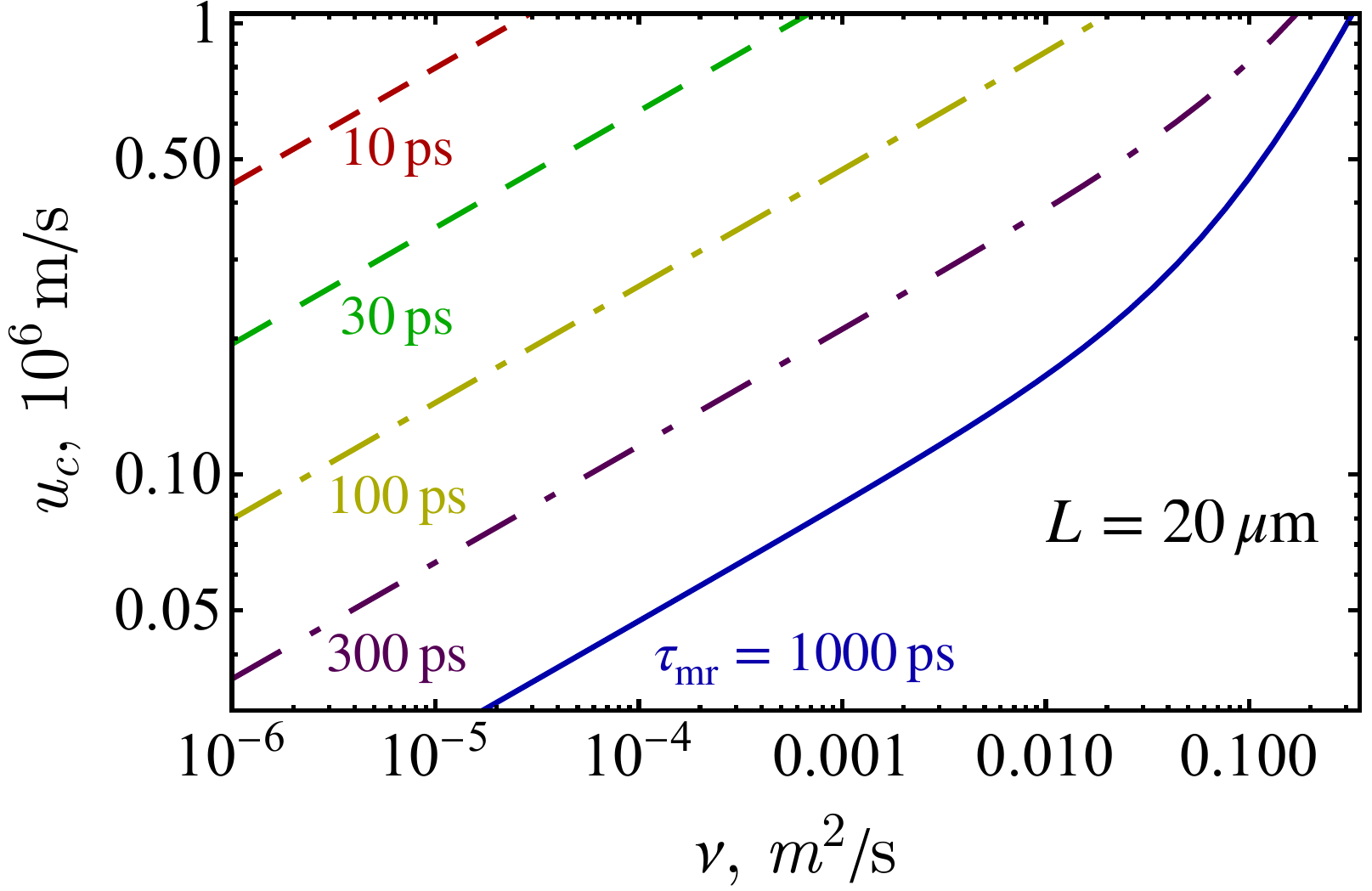} &
\includegraphics[width=0.33\linewidth, clip=true]{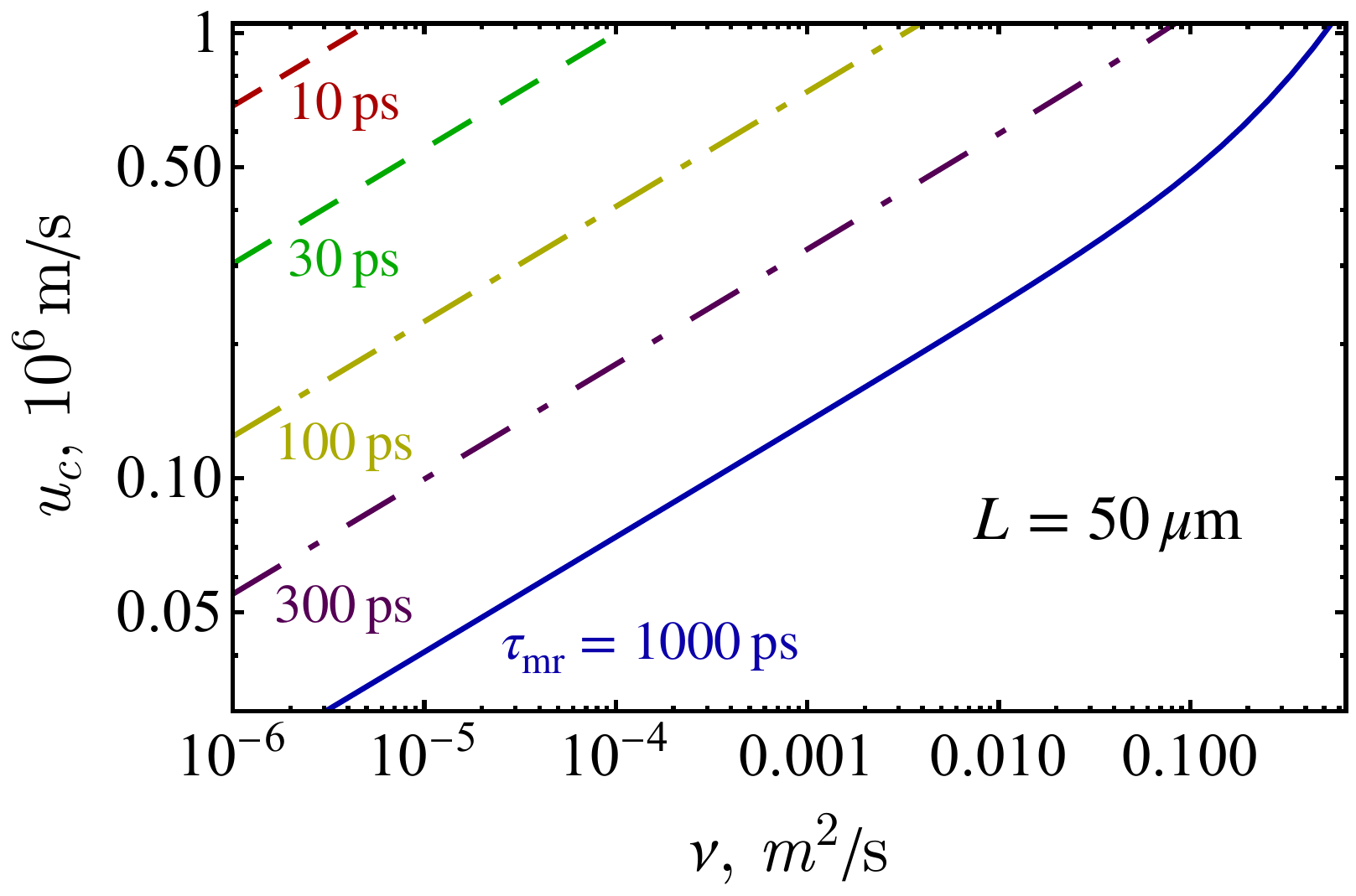} & 
\includegraphics[width=0.33\linewidth, clip=true]{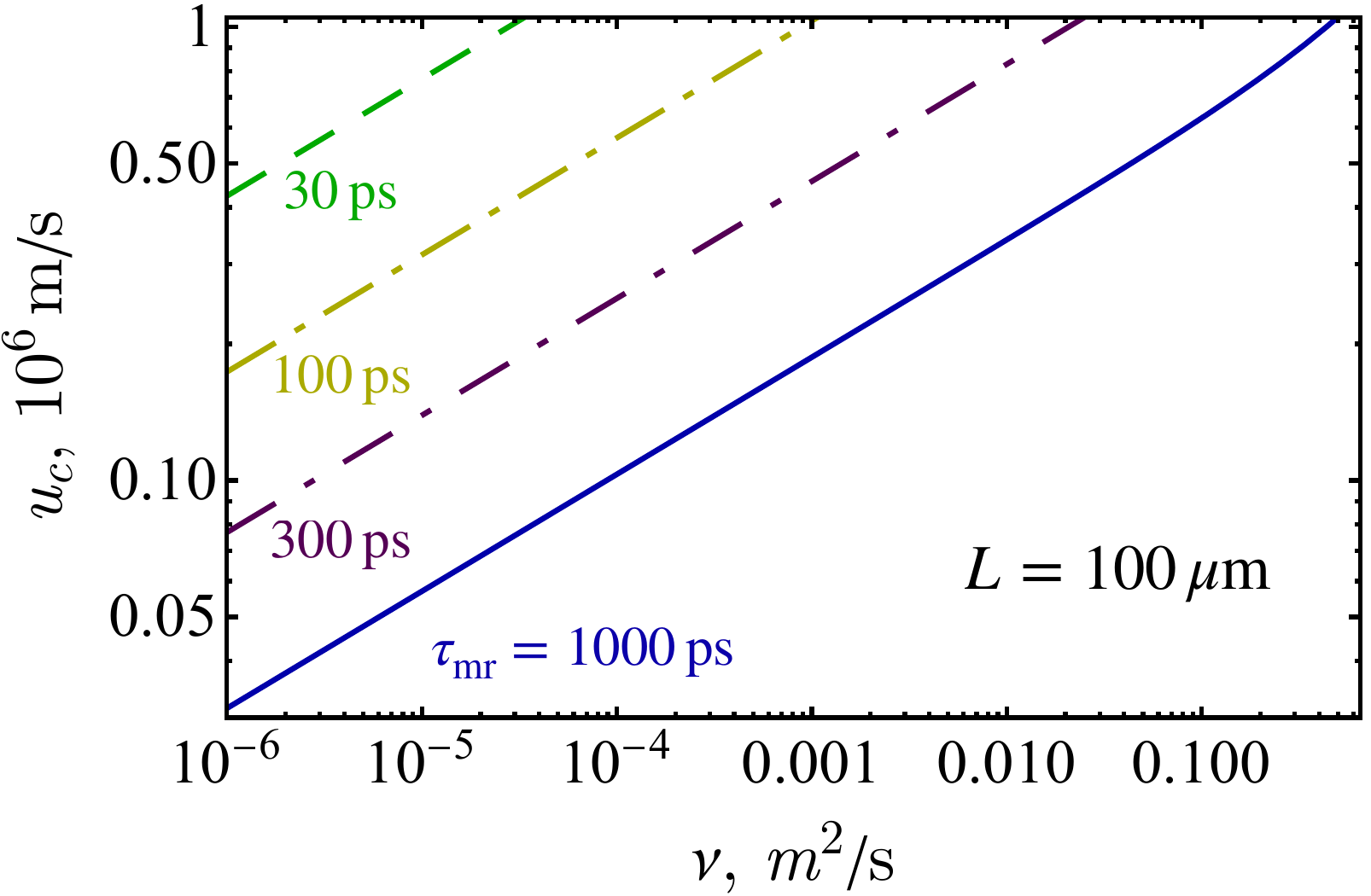} 
\end{tabular}
\end{center}
\caption{The critical velocity $u_c$ of the initial electronic flow vs. viscosity $\nu$ at various relaxation times $\tau_{\mr}$ for the circular hole of various diameters $L = 0.1, \dots, 100\,\mu$m as derived from the fractional power law~\eq{eq_tau_c} which is assumed to be valid in the whole range of parameters used. The straight segments mark the validity of the asymptotic fractional-power-low regime~\eq{eq_u_c_lim} which is realized at short relaxation times $\tau_{\mr}$ and characterized by the scaling $u_c(\tau_{\mr}) \propto \nu^{\frac{\alpha - 1}{\alpha}} = \nu^\gamma$ with the fixed exponent $\gamma \equiv 0.258(1) \simeq 1/4$. The electronic fluid enters the preturbulence regime at $u > u_c$.}
\label{fig_velocity}
\end{figure*}

\clearpage

\section{D. Frequencies and magnitude of velocity oscillations for various sizes of holes.}

\begin{figure*}[!htb]
\begin{center}
\includegraphics[width=0.95\linewidth, clip=true]{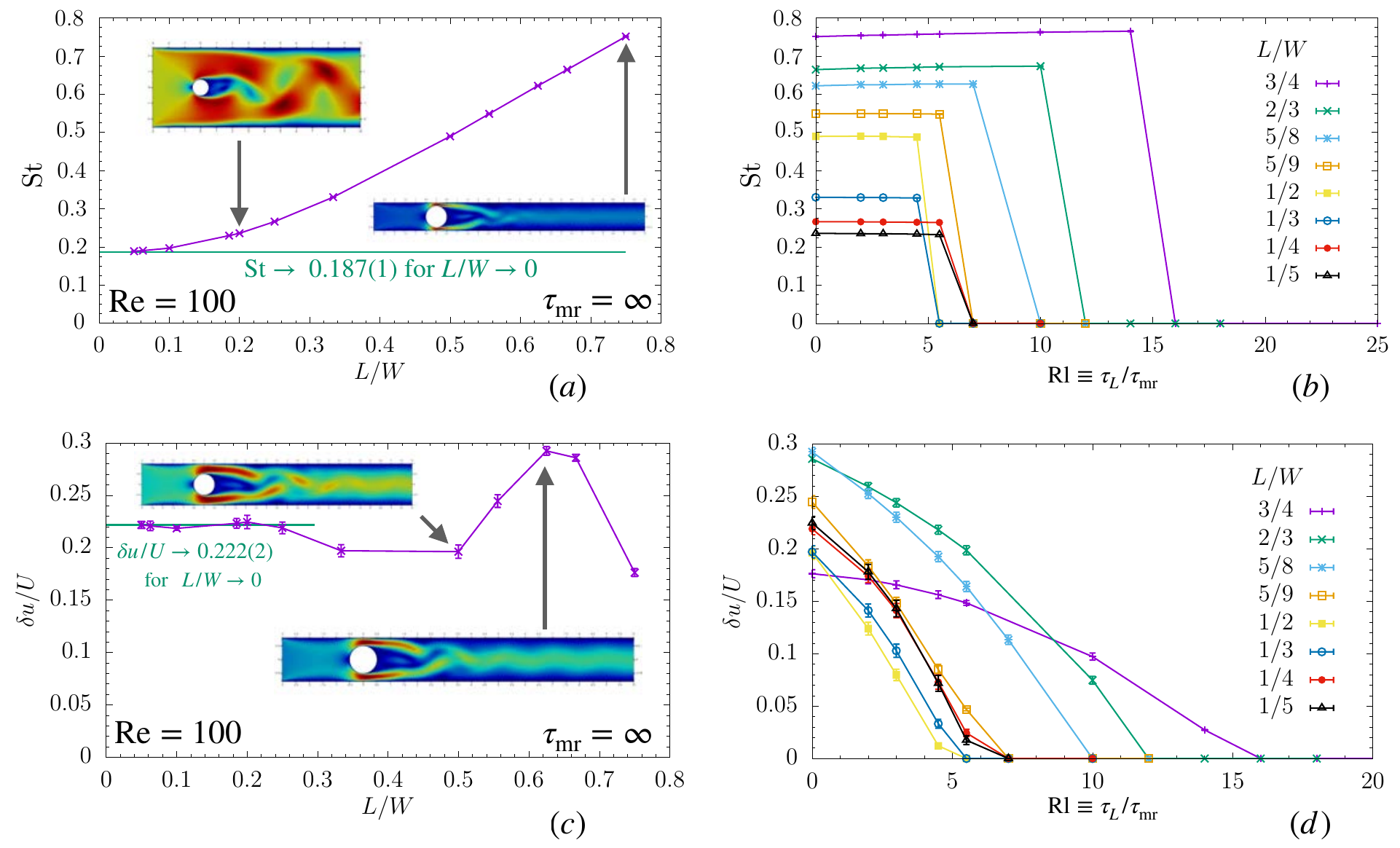}
\end{center}
\caption{The Strouhal number~\eq{eq_f_St} of flow oscillations at $\re = 100$ (a) in the dissipationless limit, $\tau_{\mr} = \infty$, as a function of the ratio $L/W$ of the hole diameter $L$ and the width of the channel $W$; (b) as a function of the relaxation number $\rl$ for various ratios $L/W$. (c) and (d): The same as in (a) and (b), but for the magnitude of the velocity oscillations at the point at a distance $3 L$ past the hole (cf. Fig.~\ref{fig_large_hole} in the main text). The frequency does depend on the geometry (a) but is largely independent of the relaxation time $\tau_{\mr}$ (b). The oscillations of the flow velocity exhibit substantial dependence on both parameters. The insets show examples of the fluid flow at particular geometries and also the limits at a vanishing diameter of the hole.}
\label{fig_varying_diameters}
\end{figure*}

\end{document}